\documentclass[prb,twocolumn,showpacs,preprintnumbers,amsmath,amssymb]{revtex4}

\usepackage{graphicx}
\usepackage{dcolumn}
\usepackage{bm}

\newcommand{\cddyz}{Cd-$d_{yz}$}
\newcommand{\opy}{O1-$p_y$}
\newcommand{\opz}{O1-$p_z$}
\newcommand{\neutra}{\mbox{72A-SC(0)}}
\newcommand{\carg}  {\mbox{72A-SC(-2)}}
\newcommand{\fluor} {\mbox{72A-SC(F)}}
\newcommand{\unitefg}{\mbox{10$^{21}$V/m$^2$}}
\newcommand{\cgav}{V$_{33}$}

\begin{document}
\flushbottom

\title{\bf The problem of a metal impurity in an oxide:  {\it ab initio} study of
electronic and structural properties of Cd in Rutile TiO$_2$}

\author{Leonardo A. Errico}
\altaffiliation[Post-Doctoral Fellow of ]{Consejo Nacional de
Investigaciones Cient\'{\i}ficas y T\'ecnicas (CONICET),
Argentina} \email{errico@fisica.unlp.edu.ar}
\author{Gabriel Fabricius$^\dag$}
\author{Mario Renter\'{\i}a}
\altaffiliation[Member of ]{CONICET}
 \email{fabriciu@fisica.unlp.edu.ar,renteria@fisica.unlp.edu.ar}

 \affiliation{ Departamento de
F\'{\i}sica, Facultad de Ciencias Exactas, Universidad Nacional de
La Plata, C.C. N$^\circ$67, 1900 La Plata, Argentina}

\date{\today}

\begin{abstract}
In this work we undertake the problem of a transition metal
impurity in an oxide. We present an {\it ab initio} study of the
relaxations introduced in TiO$_2$ when a Cd impurity replaces
substitutionally a Ti atom. Using the Full-Potential
Linearized-Augmented-Plane-Wave method we obtain relaxed
structures for different charge states of the impurity and
computed the electric-field gradients (EFGs) at the Cd site. We
find that EFGs, and also relaxations, are dependent on the charge
state of the impurity. This dependence is very remarkable in the
case of the EFG and is explained analyzing the electronic
structure of the studied system.  We predict fairly anisotropic
relaxations for the nearest oxygen neighbors of the Cd impurity.
The experimental confirmation of this prediction and a brief
report of these calculations have recently been presented [P.R.L.
89, 55503 (2002)]. Our results for relaxations and EFGs are in
clear contradiction with previous studies of this system that
assumed isotropic relaxations and point out that no simple model
is viable to describe relaxations and the EFG at Cd in TiO$_2$
even approximately.
\end{abstract}

 \maketitle

\section{\label{sec:1}INTRODUCTION}
The problem of metal impurities in oxides is a challenge in
solid-state physics both from a fundamental and applied point of
view. Description of impurity levels in oxide semiconductors has a
great technological interest, but a complete theoretical
description of the problem is very difficult. In effect,
impurities introduce atomic relaxations in the host and modify the
electronic structure of the system, being the interplay between
these two effects one of the principal difficulties in the
theoretical approach. Some experimental techniques used in
material science introduce probe-atoms into the systems to be
studied,\cite{techniques1,techniques2}  giving valuable physical
information {\it seen} by the probe-atom, usually an impurity in
the system. In particular, TDPAC spectroscopy~\cite{TDPAC} is a
hyperfine technique that enables a high resolution  determination
of the EFG tensor at impurity sites. Due to the r$^{-3}$
dependence of the EFG operator from the charge sources, the EFG is
mostly originated in the electronic charge density close to the
impurity nucleus, which in turn reflects the probe chemistry with
its neighborhood. Thus, it would be important that the theoretical
approach to the metal-impurity problem would be
able to  predict the EFG tensor with the required accuracy. Since
the EFG is very sensitive to the anisotropic charge distribution
close to the nucleus, for its accurate calculation the entire
electronic configuration of the host, perturbed by the presence of
the impurity, has to be determined. This kind of calculations
should be performed  at the {\it ab initio} level and can be done
in the frame of the Density Functional Theory (DFT). A well proven
method to solve the all-electron DFT equations in solids is the
Full-Potential Linearized-Augmented-Plane-Wave (FLAPW)
method.\cite{flapw} This method was able to predict with high
accuracy and precision the EFG at undisturbed lattice sites in a
large variety of pure systems.\cite{blaha,efg} However, on systems
with impurities, very few calculations have been performed and the
state-of-the-art is far from being routinely in this field.
\cite{state-of-the-art,state2,state3,
state4,state5,state6,state7,state8,state9} The first fully
self-consistent {\it ab initio} determination of the EFG tensor at
an impurity site (Cd) in an oxide (TiO$_2$) has recently been
reported~\cite{prl}. In that work we performed a
 FLAPW calculation of the relaxations introduced by
the impurity and studied their interplay with the electronic
structure of the system, predicting highly anisotropic relaxations
of the nearest neighbors of the impurity and a drastic change in
the orientation of the principal component of the  EFG tensor.
This prediction was confirmed in the same work by a key PAC
experiment. Due to the complexity of the physical situation we
want to describe, even in the framework of {\it ab initio}
calculations some variables have to be determined through
comparison with experiments. For example, we performed the
calculations for two charge states of the impurity and one of them
could be discarded because it gave a value for the EFG asymmetry
parameter, $\eta$, incompatible with experience.

The central purpose of this work is to present and discuss the
electronic structure of the TiO{$_2$}:Cd impurity system with more
detail that we could do in Ref.~\onlinecite{prl}. Here, we study
the relationship between the calculated EFG at the impurity site -
the key experimental quantity - and the interplaying physical
quantities: the structural relaxations and the character and
filling of the impurity state. We also discuss in detail the
precision of our calculations studying  its dependence on the
different parameters that control the precision (impurity
dilution, base dimension, average in k-space, etc.)
 and its dependence on the approximation used for
the exchange-correlation potential.

Another important question to be addressed concerning all
experimental techniques that introduce impurity-tracers in solids
is at what extent simple models usually used to predict the
measured quantities are acceptable in view of the modifications
introduced by the presence of the impurity. Concerning the EFG,
simple approximations like the point-charge
model~\cite{techniques2} with antishielding
factors\cite{stern,lieb} and isotropic relaxations of
nearest-neighbor atoms~\cite{lieb,akai} are commonly used. In this
work we give arguments against the applicability of these simple
models for the studied system.

This paper is organized as follows. In section II we describe the
method of calculation used in the present work and include
 some preliminary calculations for an unrelaxed
 cell in order to discuss methodologically
  the point of the charge state of the impurity.
  In section III.(A-C) we present and discuss the results of performing
  relaxations of the Cd nearest-neighbors (NN) in a 72-atoms super-cell for two different states of charge of
  the impurity, while in section III.D we study the accuracy of
  these results performing several additional calculations.
  Even if section III.D
  could be found rather {\it technical} by people not specially
  interested in first-principles calculations, we think it will
  be useful for the {\it ab initio} community
  interested in the problem of impurities in solids.
  In section III.E we compare our results with previous studies
  of this system. Finally, in section IV we present our conclusions.

\section{\label{sec:2}METHOD OF CALCULATION}

\subsection{\label{sec:2.A}A general scheme}

        Our aim is to calculate from first principles
the structural relaxations produced in the TiO$_2$ lattice when a
Cd impurity replaces a Ti atom and the electronic structure of the
resulting system. In particular we are interested in the EFG
tensor at the Cd site. The general scheme we adopted to deal with
this problem was the following: we considered a super-cell (SC),
containing a single Cd impurity, repeated periodically and
performed first-principles calculations  We studied the relaxation
introduced by the impurity computing the forces on the Cd
neighbors and moving them until the forces vanished. The
calculations were performed with the  {\sc wien97}
 implementation, developed by
 Blaha {\it et al.},\cite{wien97} of  the Full-Potential
 Linearized-Augmented-Plane-Wave (FLAPW) method and
we worked in the LDA approximation using the Perdew and Wang
parametrization for the exchange-correlation potential.\cite{LDA}
 In this method the unit cell is divided into
non-overlapping spheres with radii R$_i$ and an interstitial
region. The atomic spheres radii used for Cd, Ti and O were 1.11,
0.95 and 0.85~{\AA}, respectively. We took for the parameter
RK$_{MAX}$, which controls the size of the basis-set in these
calculations, the value of 6 that gives basis-set consisting in
4500 LAPW functions for the SC described in section~\ref{sec:2.B}.
We also introduced local orbitals (LO) to include Ti-3$s$ and
3$p$, O-2$s$ and Cd-4$p$ orbitals.
 Integration in
reciprocal space was performed using the tetrahedron method taking
an adequate number of {\it k}-points in the first Brillouin zone.
Once self-consistency of the potential was achieved,
quantum-mechanical-derived forces were obtained according to Yu
{\it et al.}\cite{forces}, the ions were displaced according to a
Newton damped scheme \cite{newton} and new positions for Cd
neighbors were obtained. The procedure is repeated until the
forces on the ions are below a tolerance value taken as
\mbox{0.025 eV/\AA}. At the relaxed structure, the $V_{ij}$
elements of the EFG tensor are obtained directly from the $V_{2M}$
components of
 the lattice
harmonic expansion of the self-consistent potential (a more
detailed description of the formalism used to compute the EFG
tensor may be found in Ref.~\onlinecite{efg}).

There is still an important point to take into account
concerning the calculation of the electronic structure
of an impurity system: the charge state of the impurity. We
let this point to the end of this section as it will
be easily handled after the analysis of some preliminary
calculations.

\subsection{\label{sec:2.B} Cell and super-cell}

\begin{figure}[h]
\includegraphics*[bb= 3cm 13.9cm 12cm 28.7cm, viewport=0cm 0.5cm 9cm 14cm, scale=0.8]{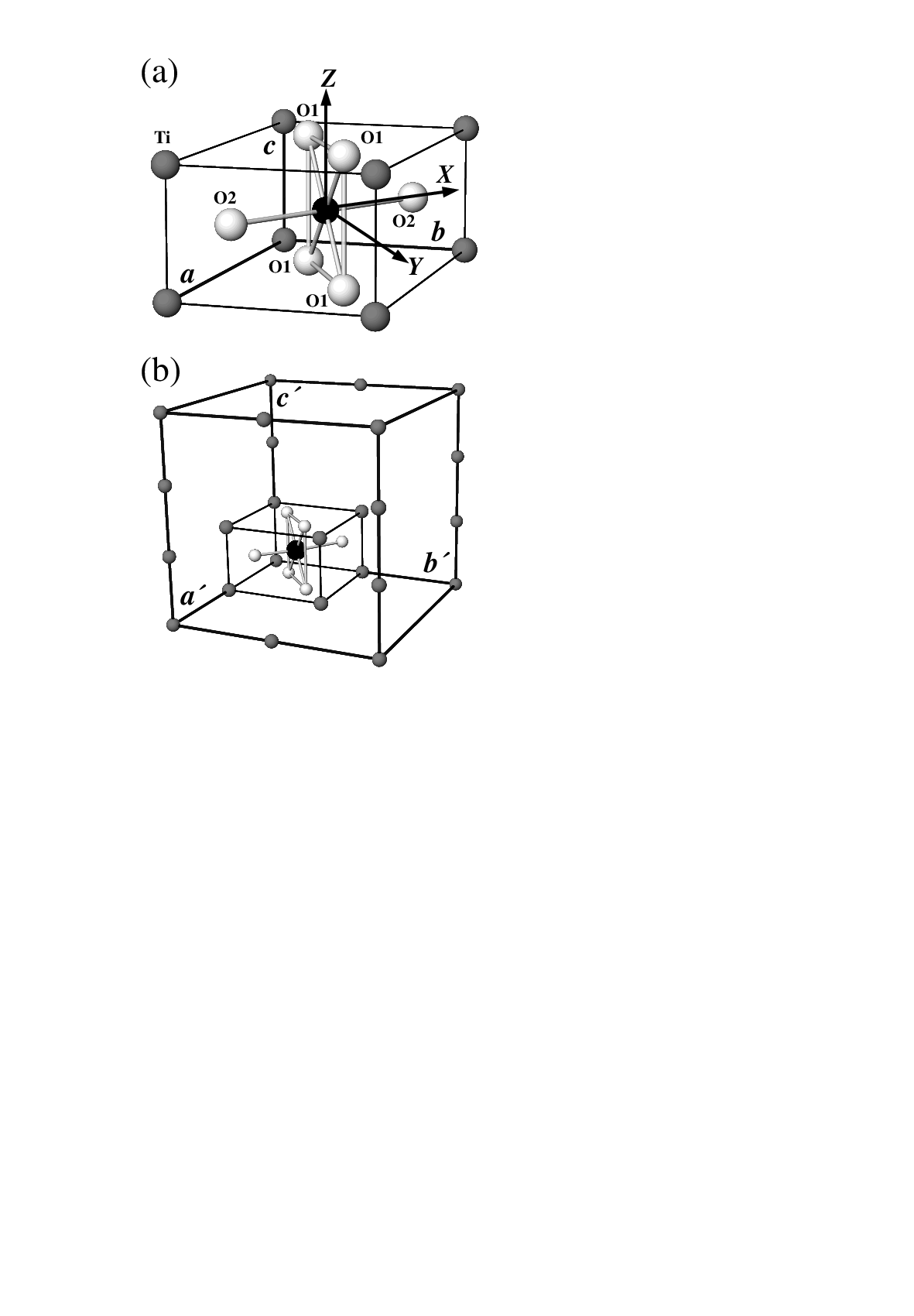}
\caption{\label{cell} (a) Unit cell of rutile TiO$_2$. All the
results discussed in this article are referred to the axis system
 indicated in this figure assuming that Cd replaces the black Ti
atom. Even when O1 and O2 atoms are equivalent in TiO$_2$, it will
not be the case in the impurity system; (b) 72-atoms super-cell
used in our calculations. Some O and Ti atoms are not shown for
clearness.}
\end{figure}

 The rutile (TiO$_2$) structure is tetragonal
 ($a=b=$ 4.5845$_{1}$~{\AA}, $c=$ 2.9533$_{1}$~{\AA}). The
unit cell (shown in Fig.~\ref{cell}) contains 2 metal atoms (Ti)
at positions 2$a$ (0, 0, 0) and (1/2, 1/2, 1/2) and 4 anions (O)
at positions 4$f$ $\pm$(u, u, 0;  u+1/2, 1/2-u, 1/2), with u =
0.30493$_7$.\cite{tio2}  The super-cell considered in the present
work
  consists of twelve unit cells of TiO$_2$ where one Ti atom has been
  replaced by a Cd atom. The resulting 72-atom SC
  (called 72A-SC in the future) has dimensions
  $a'=$2$a$, $b'=$2$b$, $c'=$3$c$ and is also tetragonal with
   $c'/a'=$0.97 giving an almost cubic lattice. This SC
  keeps Cd atoms as far as possible from each other for the given
  cell volume. For checking
  purposes we have also considered SCs containing
  eight and sixteen unit cells.
  We assume that relaxations
  to be performed preserve point group symmetry of the cell
  in this initial configuration.
  Symmetry restricts O1 and O2 displacements to $yz$ plane and $x$
direction respectively. In order to check this assumption
  and the stability of the obtained solution, at the end of the
  relaxation process we performed new calculations with O1 and O2
  atoms displaced from their symmetry positions and verified that
  these solutions are a minimum for each system studied.

   The self-consistent calculations were
performed taking 8 $k$-points in the irreducible Brillouin zone
for the metallic system (situation $(i)$), and 2 $k$-points for
the non-metallic ones (situation $(ii)$). In order to plot the DOS
we calculate eigenvalues at a denser mesh of 36 $k$-points.

\subsection{\label{sec:2.C} Preliminary study: Charge state of the impurity}

\begin{figure}[!]
\includegraphics*[bb= 0.3cm 0.3cm 5.9cm 5.2cm, viewport=0.1cm 0.1cm 8cm 5cm, scale=1.5]{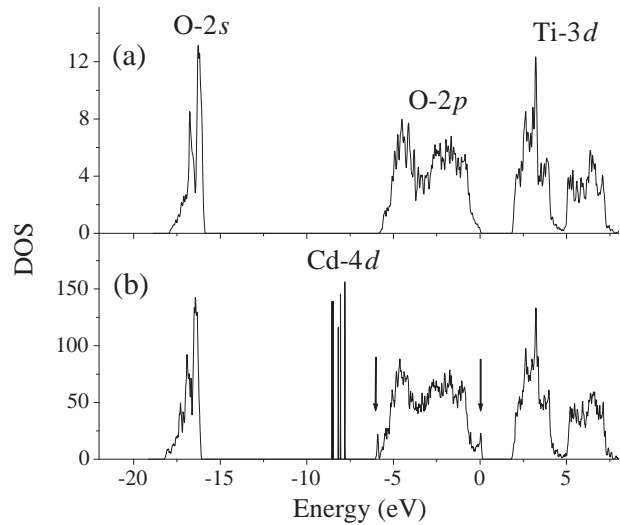}
\caption{\label{DOS1}  (a) DOS for pure TiO$_2$. (b) DOS for the
72-atom super-cell (unrelaxed). Energies are referred to the Fermi
level. Note that the Cd-$d$ band is described as very sharp peaks
indicating that interaction between Cd atoms of different cells is
quite small for the considered super-cell. The arrows indicate
impurity states in the valence band.}
\end{figure}

\begin{figure}[h]
\includegraphics*[bb= 3.5cm 18.7cm 10.4cm 27cm, viewport=0cm -0.2cm 7cm 8.5cm, scale=1.1]{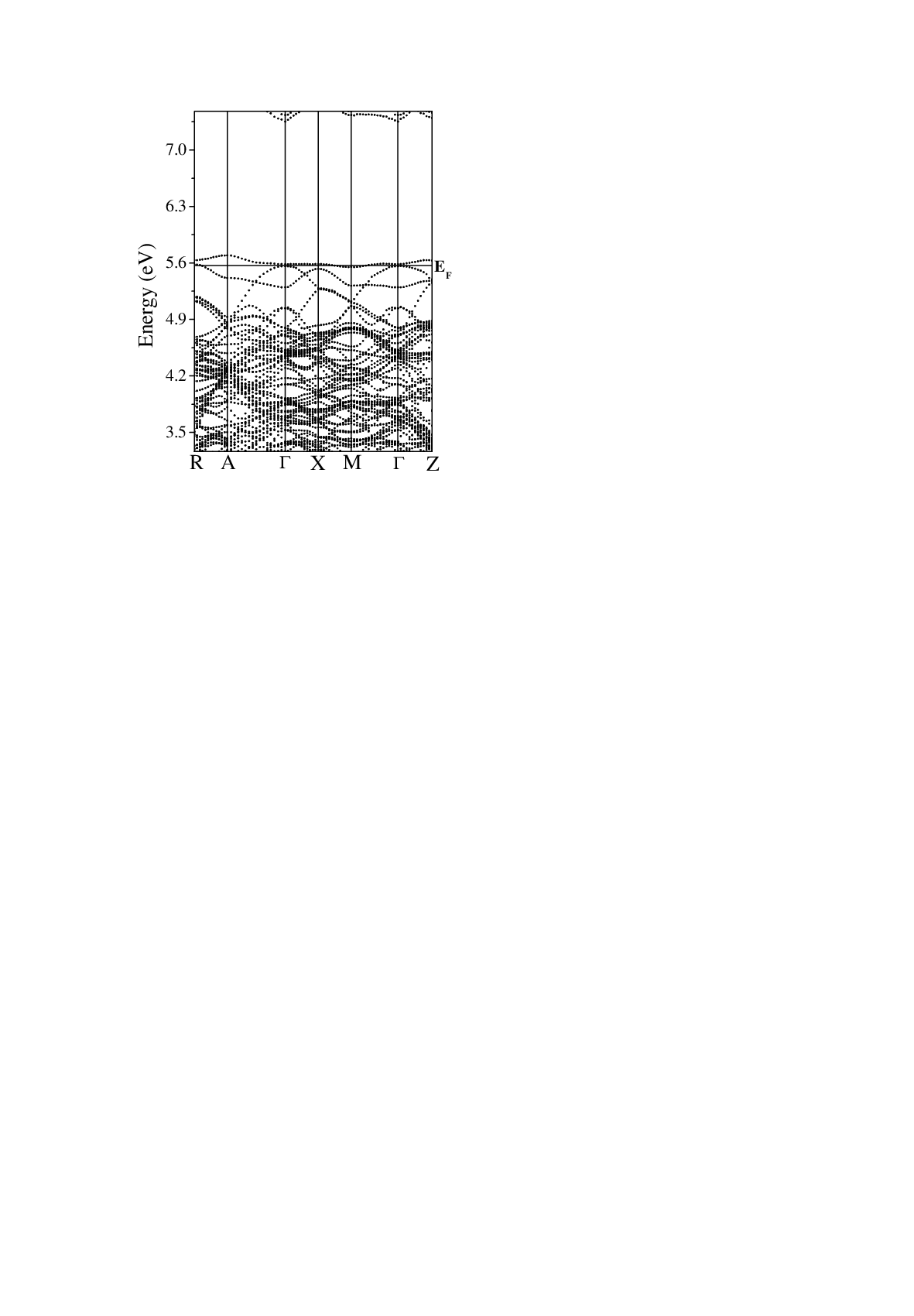}
\caption{\label{bands_c12n-sr}Band structure for the 72-atom
super-cell (unrelaxed). The zero of energy is defined as the
average Coulomb potential in the interstitial region.}

\end{figure}

In order to discuss the problem of the charge state of the
impurity we have done some preliminary calculations. We have
calculated the self-consistent electronic structure of the 72A-SC
 with all atoms in their initial unrelaxed positions.
 We want to analyze the changes in the electronic structure
 of the system caused by the presence of the impurity,
 neglecting for the moment the problem of structural
 relaxations.
The density of states of this system is compared with the one of
pure TiO$_2$ in Fig.~\ref{DOS1}.
Pure Ti$^{+4}$O$_2 ^{-2}$ is a wide band-gap semiconductor with
the O-$p$ band filled and the Ti-$d$ band empty. Since Cd valence
is 2+, when a Cd atom replaces a Ti atom in the SC the resulting
system is metallic because of the lack of two electrons necessary
to fill up the O-$p$ band. The question that arises at this point
is if the real system we want to describe provides the lacking two
electrons (via an oxygen vacancy, for example) or not, and if this
point is relevant for the present calculation. We will show first
that this point is absolutely relevant for the present
calculation. Comparison of Fig.~\ref{DOS1}(a) and (b) shows that
the presence of the Cd impurity in the SC introduces  the
appearance of
 Cd-$d$ levels and impurity states at the top and the bottom
 of the valence band in the corresponding
DOS. The ones at the top of the valence band can be better seen
looking at the band structure in this energy range shown
 in Fig.~\ref{bands_c12n-sr}.
The two bands that are immediately above and below the Fermi level
when crossing A-point correspond
  to impurity anti-bonding states that are spatially located
 at Cd and at their O1 and O2 nearest-neighbor atoms. In particular,
the wave function of the impurity state that
 remains almost completely unoccupied has
 character \cddyz,
 \opy, and \opz. Then, providing two electrons
 to the system implies a drastic change in the symmetry
of the  electronic charge distribution
  in the neighborhood of the impurity. Therefore, different
  nearest-neighbors relaxations and also different
 EFG could be expected for different charge states of the impurity.

In the present work, we present calculations for two charge states
of the impurity system, corresponding to two different physical
situations: $(i)$ Cd$^0$ (neutral impurity state), corresponding,
e.g., to an extremely pure crystal at low temperatures, and $(ii)$
Cd$^-$ (charged acceptor state), the system provides two electrons
via an oxygen vacancy, donor defects, etc. To study case $(i)$ we
used the 72A-SC already described, to study case $(ii)$ we also
used the 72A-SC but performed self-consistent calculations adding
two electrons to the
 SC that we compensate with an homogeneous positive
 background in order to have a neutral cell to
 compute total energy and forces
 (this procedure is implemented since version {\sc wien97.9}
 of the FLAPW package). We have also simulated situation $(ii)$
with an alternative procedure: we replaced in the 72A-SC the two
most distant oxygen atoms by two fluorine atoms. In this
 way we provide two electrons to fill up the O-$p$ band
  without introducing any artificial background. We expected that
  the difference between flourine and oxygen
  potentials should modify only slightly the results.
  In summary, we performed self-consistent FLAPW calculations
  for the following systems:
\begin{tabbing}
 $ (i)$ \hskip 0.3cm \=  Cd$^0$, \hskip 1.8cm \= (TiO$_2$)$_{23}$CdO$_2$ \\
 $ (ii)$ \>  Cd$^-$(2e), \> (TiO$_2$)$_{23}$CdO$_2 + 2e^-$ \\
      \> Cd$^-$(fluorine), \>      (TiO$_2$)$_{23}$CdF$_2$
\end{tabbing}

\section{RESULTS AND DISCUSSION}

\subsection{\label{sec:3.A} Structural relaxations}

Let us first consider the relaxation of only the 6 nearest oxygen
neighbors of the Cd impurity (O1 and O2 in Fig~\ref{cell}.) In
Table~\ref{tab:distances} we compared the results of the
relaxation of these oxygen atoms  for the different systems
studied. We see that for both charge-states of the impurity the
relaxations are quite anisotropic, with the Cd-O1 distance larger
than Cd-O2 distance, opposite to the initial unrelaxed structure.
This result is opposite to what other authors have assumed in
previous studies of this system \cite{akai,lieb} and confirms the
tendency predicted in our previous calculation  with a much
smaller SC.\cite{our_nqi}  As it can be seen in
Table~\ref{tab:distances} the difference in the charge state of
the impurity affects essentially the relaxation of O1 atoms that
present a slightly larger relaxation for the charged impurity.
Relaxation for the two ways of simulating the charged state of the
impurity (case Cd$^-$) are very similar indicating that both
approaches are well suited to deal with this problem.

\begin{table}
\caption{\label{tab:distances}Final coordinates of the Cd nearest
oxygen neighbors for the different calculations performed with the
72A-SC compared with the ones of pure TiO$_2$. $d$(Cd-O1) and
$d$(Cd-O2) are the distances (in \AA) from Cd to O1 and O2 atoms,
respectively. $\theta _1$ is the angle (in degrees) between $z$
axis and
 $\vec r_{O1} - \vec r_{Cd}$.}
\begin{ruledtabular}
\begin{tabular}{lcccc}
           & & $d$(Cd-O1)  & $d$(Cd-O2) & $\theta _1$  \\
\hline
 TiO$_2$   & &  1.944    &   1.977  &  40.47   \\
 Cd$^0$   & &  2.153    &   2.108  &  39.59   \\
 Cd$^-$(2e)     & &  2.185    &   2.111  &  39.55   \\
 Cd$^-$(fluorine)    & &  2.191    &   2.121  &  39.73    \\
\end{tabular}
\end{ruledtabular}
\end{table}

\begin{figure}[h]
\includegraphics*[bb= 2.2cm 19.1cm 11.1cm 22.5cm, viewport=-0.2cm 0.3cm 10cm 4cm, scale=0.9]{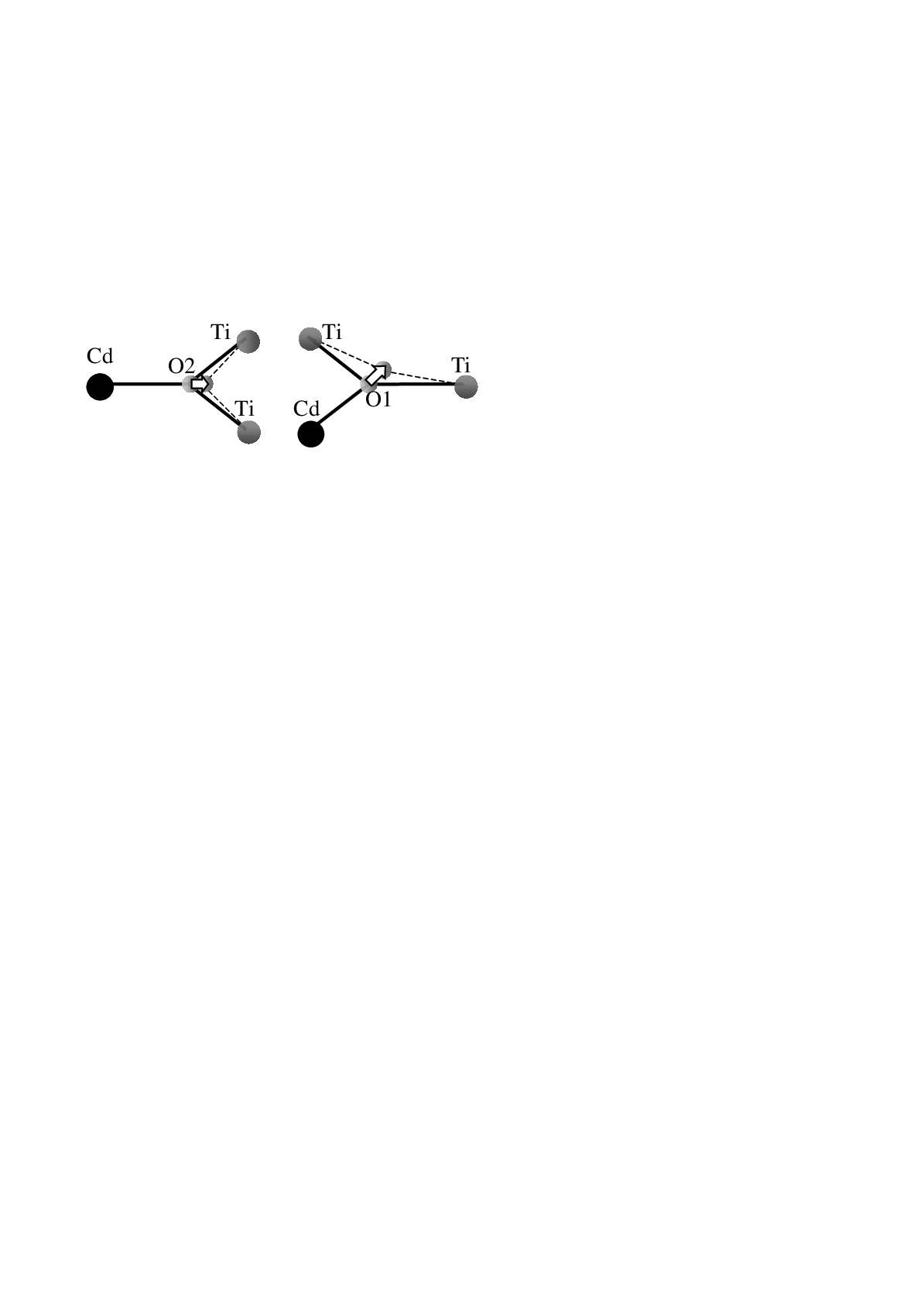}
\caption{\label{stretching}
 Planes {\bf XZ} and {\bf YZ} (see Fig.~\ref{cell}) containing O2 and O1
 atoms, respectively, with their neighbors. The arrows indicate the
displacement of the oxygen atoms from the unrelaxed  to the final
relaxed positions in the 72A-SC[Cd$^-$(2e)] system. The size of
the relaxation has been duplicated in order to better visualize
the effect.}
\end{figure}

Anisotropy in the relaxations of the nearest oxygen neighbors of
the Cd impurity can be understood by inspection of
Fig.~\ref{stretching}.
Stretching of Cd-O2 bond implies a considerable shortening in
Ti-O2 bonds. However, stretching of Cd-O1 bond doesn't affect so
much Ti-O1 bonds since the structure is more open in this
direction. So, at the end of the relaxation, Cd-O1 bond stretches
almost twice than Cd-O2 bond.

\subsection{\label{sec:3.B} Electronic structure}

\begin{figure*}[t]
\includegraphics*[bb= 2.7cm 15.3cm 18.7cm 23.6cm, viewport=-1cm -0.1cm 20cm 9cm, scale=1]{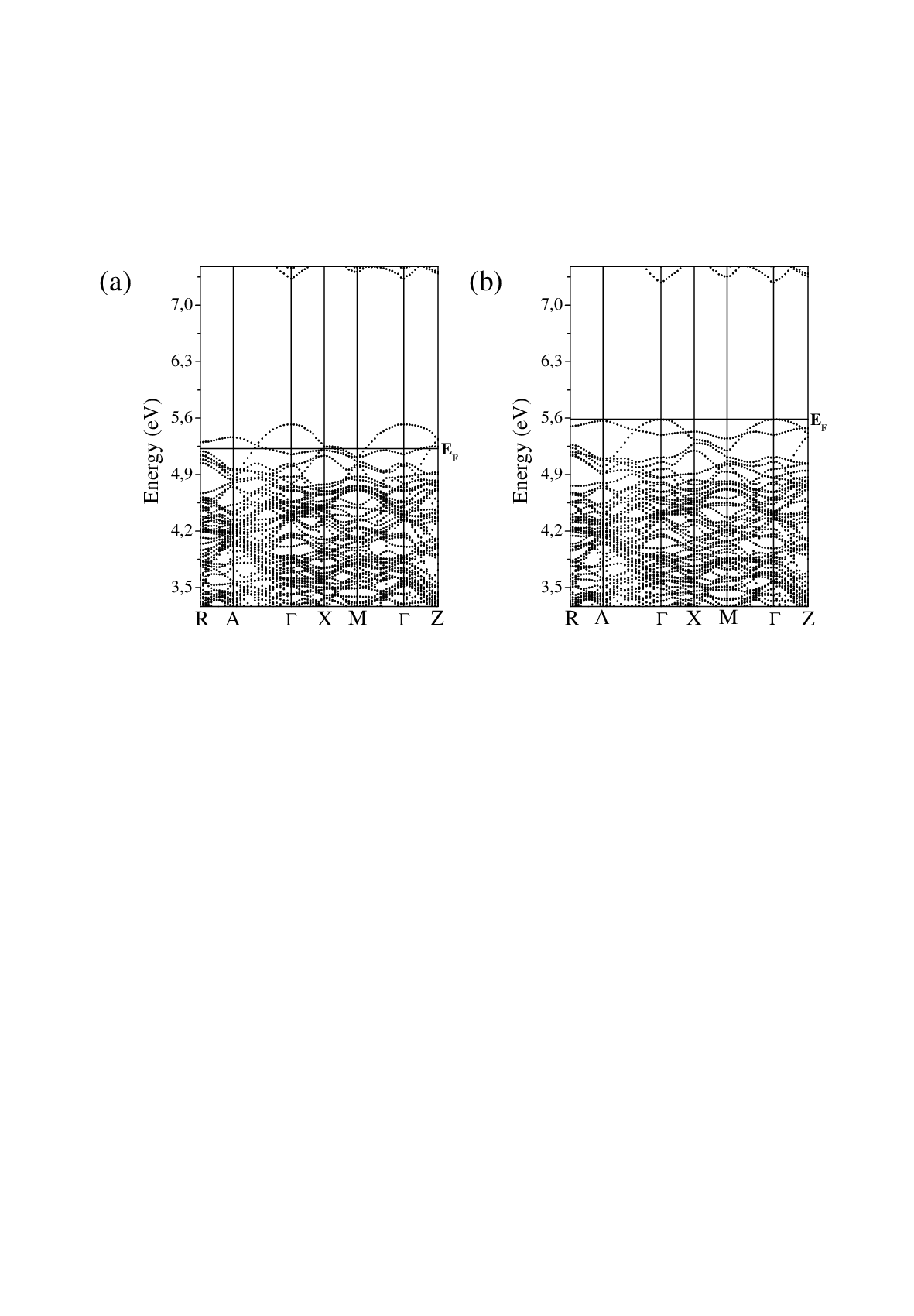}
\caption{\label{bands}Band structure of the relaxed 72A-SC for
different charge states of the impurity: (a) Cd$^0$ (neutral
state) and (b) Cd$^-$(2e) (charged state). The zero of energy as
in Fig. 3.}
\end{figure*}

\begin{figure}[!]
\includegraphics*[bb= 3.4cm 19cm 8.9cm 26.9cm, viewport=0cm 1.1cm 9cm 8cm, scale=1.5]{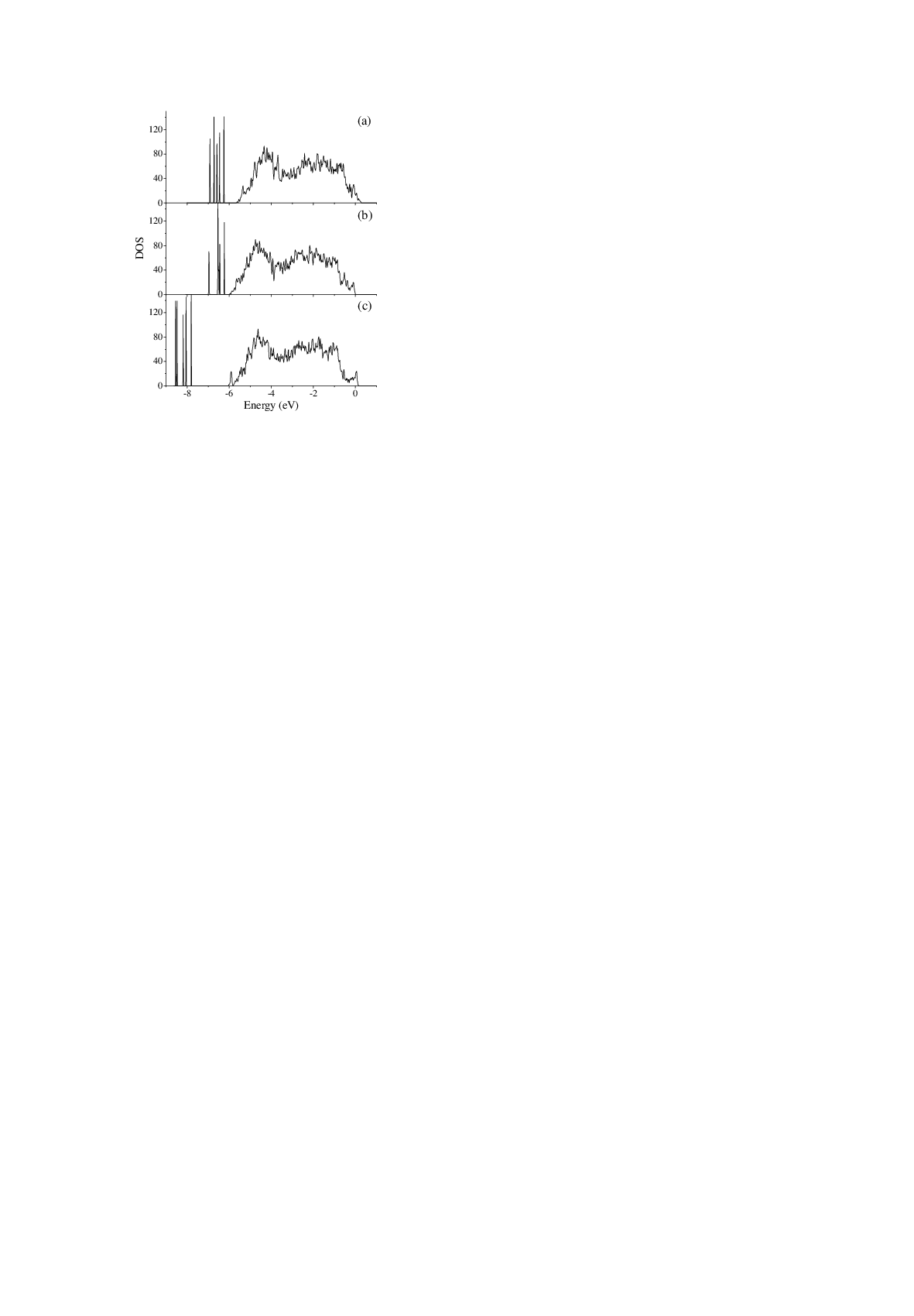}
\caption{\label{DOS2} DOS of the (a) relaxed 72A-SC (Cd$^0$), (b)
relaxed 72A-SC[Cd$^-$(2e)], and (c) unrelaxed 72A-SC (Cd$^0$).
Figure (c) is the same as Fig.~\ref{DOS1}(b) and has been repeated
here in other scale for the sake of comparison. Energies are
referred to the Fermi level.}
\end{figure}

\begin{figure*}[!]
\includegraphics*[bb= 2cm 12cm 20.6cm 23cm, viewport=0.8cm -0.1cm 20cm 11cm, scale=1]{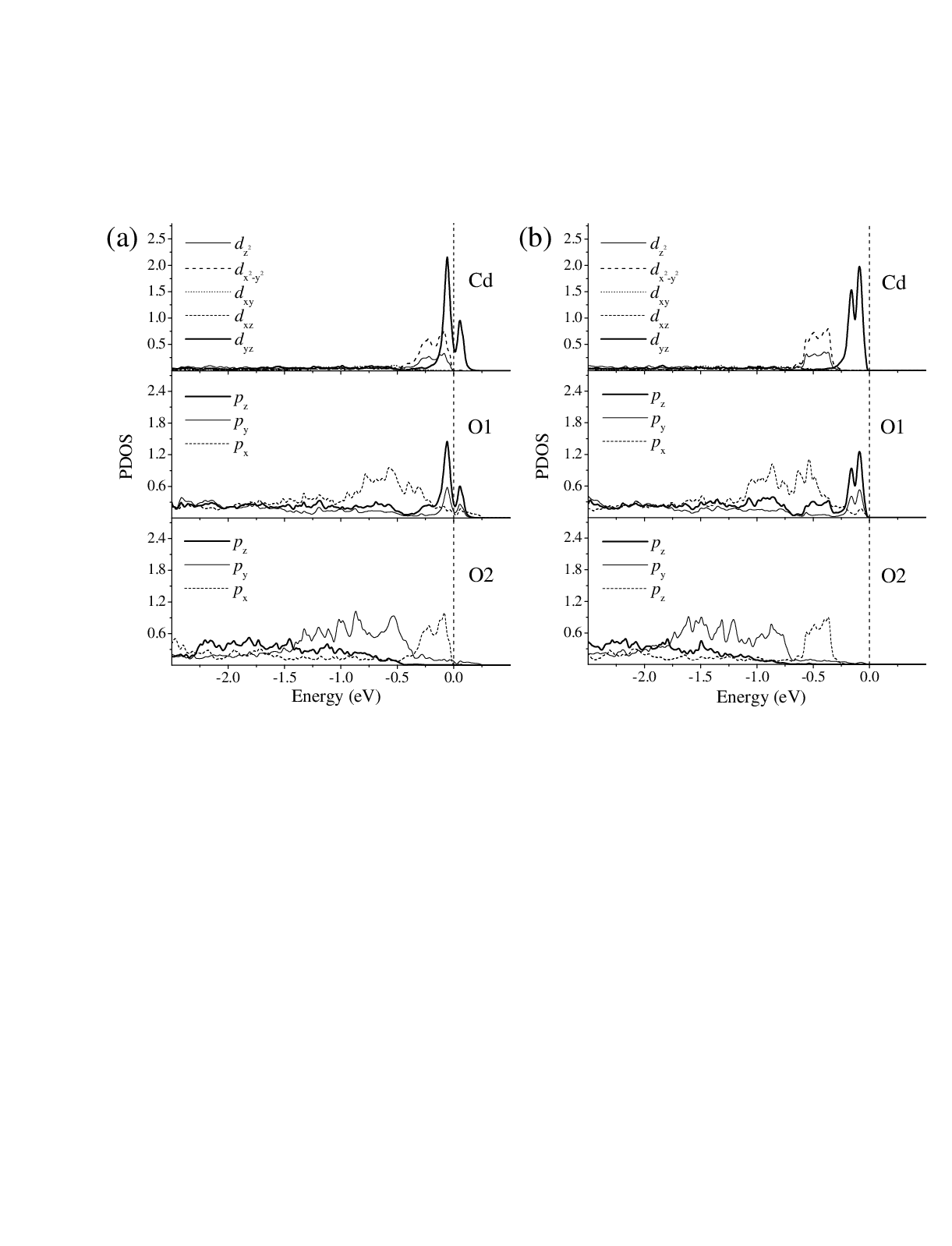}
\caption{\label{PDOSNyC} Atom-resolved PDOS at Cd, O1, and O2
atoms in the relaxed 72A-SC for (a) the neutral  charge state
(Cd$^0$) and (b) the charged
 state [Cd$^-$(2e)] of the impurity.}
\end{figure*}

In Fig.~\ref{bands}
we show the bands for the 72A-SC for the two states of charge
considered for the impurity. Comparison of Fig.~\ref{bands}(a)
with Fig.~\ref{bands_c12n-sr}
 shows that as a consequence of relaxation the outermost impurity level
 falls down in energy and goes into the O-$p$ band becoming
 half-occupied. Relaxation increases the Cd-O1 distance
 and this produces a
  softening of the Cd-$d -$O1-$p$ interaction
and the fall down in energy of the anti-bonding impurity states.
Figure~\ref{bands} shows that the band structure of the neutral
and charged relaxed structures are very similar, but the outermost
impurity state is a slightly raised in energy when it is
completely filled (case $(ii)$) as a consequence of the larger
Coulomb repulsion. Comparison of Fig.~\ref{DOS2}(a) and
Fig.~\ref{DOS2}(c) shows the little fall down in energy mentioned
for the anti-bonding impurity states an also a little rise for the
bonding ones at the bottom of the O-$p$ band. A shift upwards of
about 1.5 eV of Cd-$d$ levels from the un-relaxed \neutra\ to the
relaxed one is also present.  To look at the orbital composition
of the impurity states we plot in Fig.~\ref{PDOSNyC}(a) and (b)
the PDOS for Cd-$d$, O1-$p$ and O2-$p$ for the two charge states
of the impurity. The impurity state near the top of the valence
band has Cd-$d_{yz}$ , O1-$p_y$ and O1-$p_z$ character and, as we
mention looking at the bands, it is shifted upwards
 in energy when it is completely filled (case Cd$^-$). In
Fig.~\ref{PDOSNyC}(a)(Cd atom) and (b)(Cd atom) it  can also be
seen
 the presence of other impurity state with
components Cd-$d_{x^2-y^2}$ and
 Cd-$d_{3z^2-r^2}$
 which is completely filled in both cases.
  This impurity state is located mainly at Cd-$d_{x^2-y^2}$
 and  O2-$p_x$ but involves also contributions from
 O1-$p_z$, Cd-$d_{3z^2-r^2}$ and O3-$p_x$ (O3 is the NN of O2 atom
 in the $x$ direction). From the present calculations the outermost impurity state has an
occupation of around 1.3~e for the 72A-SC(Cd$^0$) and 2~e  for the
72A-SC(Cd$^-$). We identify this 0.7 additional electrons in this
state as the driving force that produce the slightly larger
relaxation for O1 atoms in the charged cell with respect to the
neutral cell.
We want to mention that the fact that the impurity level falls at
the Fermi energy in the Cd$^0$ case is not fortuitous. Due to
Coulomb repulsion, the impurity level would fall below the Fermi
energy if it was empty an above if it was filled being the only
self-consistent solution to be half-occupied at exactly the Fermi
energy. The occupation of the impurity level in the 72A-SC(Cd$^0$)
is therefore a constant number and fairly independent of small
fluctuations of charge in the cell.

\subsection{\label{sec:3.C} Electric-Field Gradients}

\begin{table*}
\caption{\label{tab:efgs} EFG tensor principal components at Cd
site, $V_{ii}$(in \unitefg), for the relaxed structures of the
different systems considered in our calculation compared with
experiments and the  calculation of Sato {\it et al.}
 $\eta = (V_{11}-V_{22})/V_{33}$ ($|V_{33}|>|V_{22}|>|V_{11}|$). In the last row
EFG tensor refers to Ti site in pure TiO$_2$. Q=0.83 b(Q=0.24 b)
was used to calculate \cgav\ from the experimental quadrupole
coupling constant $\nu _Q$  at $^{111}$Cd($^{49}$Ti) sites.}
\begin{ruledtabular}
\begin{tabular} {lcccccc}
                 & $V_{XX}$ & $V_{YY}$ & $V_{ZZ}$ & $V_{33}$      &$V_{33}$-direction  & $\eta$  \\ \hline
 \neutra         & -7.16    & +6.82    &  +0.34   &  -7.16  & \textbf{X}    & 0.91    \\
 \carg           & -2.87    & +4.55    &  -1.68   &  +4.55  & \textbf{Y}    & 0.26    \\
 \fluor          & -2.46    & +4.10    &  -1.63   &  +4.10  & \textbf{Y}    & 0.20    \\
 Exp.\cite{lieb} &          &          &          & 5.23(5) &   ...            & 0.18(1)  \\
 Exp.\cite{exp2} &          &          &          & 5.34(1) &   ...            & 0.18(1)  \\
Exp.
  (singlecrystal)\cite{prl}&          &          &          & 5.34(1) &  \textbf{X} or \textbf{Y}  & 0.18(1) \\
 Calc.\cite{akai}&   +1.54  & +3.56    &  -5.09   &  -5.09  & \textbf{Z}    &0.39     \\
 Exp.(pure TiO$_2$)\cite{nmr_tio2}   &     &    &    &  2.2(1)  &
\textbf{Z} &0.19(1)     \\
\end{tabular}
\end{ruledtabular}
\end{table*}

In Table~\ref{tab:efgs} we show the results for the V$_{ii}$
principal components of the EFG tensor for the three systems
studied. The resulting EFGs for the two approaches used to
simulate the charged impurity are very similar, the difference of
0.4 x \unitefg\ in components $V_{XX}$ and $V_{YY}$ is within one
could expect for the small difference found in the oxygen
positions, since EFG is very sensitive to small structural
changes. These results agree very well (in magnitude, symmetry and
orientation, see Table~\ref{tab:efgs}) with the experimental
results obtained for the EFG at Cd impurities substitutionally
located at cationic sites in rutile TiO$_2$.\cite{prl} The
difference between the EFGs obtained for the charged and neutral
cells is very remarkable: the sign, direction and absolute value
of the largest $V_{ii}$ component ($V_{33}$) are different in both
situations, and also the value of the asymmetry parameter $\eta$.
The high $\eta$ value obtained for \neutra\
 shows that the electron availability  present in
the sample leads the impurity to be in a charged state.

In order to investigate the origin of the difference in the EFG
for the two charge states of the impurity we concentrate in the
valence contribution to the EFG which originates in the asymmetry
of the valence charge distribution inside the muffin-tin sphere.
The valence contribution is usually dominant in FLAPW calculations
and can be split in the different orbital symmetries.\cite{efg} In
Table~\ref{tab:efgs comp} we show the total valence contribution
to $V_{ii}$ and its components arising from $p$ and $d$ orbital
symmetries. We see that the largest differences correspond to
$d$-components of $V_{ii}$. This difference is originated in the
filling of the impurity state at the Fermi level that has an
important component of Cd-$d_{yz}$ symmetry as can be seen in
Figs.~\ref{PDOSNyC}(a) and (b)(Cd atom). A simple analysis in
terms of partial charges \cite{dnd}
 shows that the effect of adding
 $\delta n$ electrons to an orbital $d_{yz}$
is to produce a change in $V_{ii}$ components given by: $\delta
V_{XX}$=  $I_d \delta n$, $\delta V_{YY}$= - $I_d \delta n /2$,
$\delta V_{ZZ}$= - $I_d \delta n /2$, where $I_d$ is proportional
to $\langle 1/r^3 \rangle$ for $d$ orbitals inside the muffin tin
sphere. Integration of unoccupied \cddyz\ PDOS from
Fig.~\ref{PDOSNyC}(a)(Cd atom) gives $\delta n$= 0.074. Inspection
of Table~\ref{tab:efgs comp}
 shows that the changes $\delta V_{ii}$ are
 quite well described by this estimation giving
for $I_d$ a value around 83 $\times$ 10$^{21}$V/m$^2$. Another
interesting point is the presence of $d$ contributions to $V_{ii}$
in the 72A-SC[Cd$^-$(2e)] although the Cd-$d$ band is completely
filled (see Table~\ref{tab:efgs comp}). This curious point may be
explained because radial $d$ wave functions are energy dependent,
so, electrons in anti-bonding impurity states (at the Fermi level)
contribute to $V_{ii}$ with a different $\langle 1/r^3 \rangle$
factor than the ones of the same symmetry in Cd-$d$ levels
(between 6 and 7 eV below the Fermi level). We have found an
increment of 30$\%$ in $\langle 1/r^3 \rangle$ when going from an
energy of -7~eV to the Fermi level. This dependence of $\langle
1/r^3 \rangle$ with energy was also mentioned and verified by
Blaha {\it et al.} in Ref.~\onlinecite{cu2o} for Cu in Cu$_2$O.
Finally, it has to be mentioned that the difference in $p$
contributions to $V_{ii}$ between the 72A-SC(Cd$^0$) and the
72A-SC[Cd$^-$(2e)] is not negligible and is caused by the
different positions of the Cd nearest oxygen neighbors in each
case.

\begin{table*}
\caption{\label{tab:efgs comp}{\it p} and {\it d} valence
contributions to the electric-field gradient at Cd in TiO$_2$, in
units of
         10$^{21}$V/m$^2$, for the neutral and charged states of Cd in the 72A-SC. In the last row we give the difference
         between the values corresponding to Cd$^-$(2e)  and Cd$^0$. TOT refers to the total ({\it p}+{\it d}+{\it s-d}) valence EFG. }
\begin{ruledtabular}
\begin{tabular}{lccccccccc}
    &          &    $p$   &          &          &   $d$    &
                                     &          &   TOT    &         \\
    & $V_{XX}$ & $V_{YY}$ & $V_{ZZ}$ & $V_{XX}$ & $V_{YY}$ & $V_{ZZ}$
                                     & $V_{XX}$ & $V_{YY}$ & $V_{ZZ}$  \\
                                      \hline
72A-SC(Cd$^0$) & -1.62 & +3.17   &  -1.55   &  -6.44   &  +4.29
& +2.15
                                     &  -7.97   &  +7.31   &  +0.66   \\
72A-SC[Cd$^-$(2e)]   & -2.62 & +3.55   &  -0.93   &  -0.33   &
+1.18 & -0.85
                                     &  -2.86   &  +4.58   &  -1.72   \\
diff.   & -1.00 & +0.38   &  +0.62   &  +6.11   &  -3.11   &
-3.00
                                     &  +5.13   &  -2.73   &  -2.38   \\
\end{tabular}
\end{ruledtabular}
\end{table*}

\subsection{\label{sec:3.D} Accuracy of the present study: Further relaxations
and other tests}

The main sources of error of the present study in order to compare
with experiment are the size of the SC considered, the size of the
basis, the LDA approximation used for the exchange-correlation
potential and the relaxation process that has been restricted to
the six nearest oxygen neighbors of the Cd atom. To check the
accuracy of the present study we have performed several additional
calculations.

\subsubsection{Charged impurity: Cd$^-$}

 We first focus on the case of the
charged impurity where we have performed the most intensive tests.

\paragraph{Charge of the cell}

The differences found between the results for the
72A-SC[Cd$^-$(2e)] and 72A-SC[Cd$^-$(fluorine)]
(Tables~\ref{tab:distances} and \ref{tab:efgs}) could be taken as
a measure of the error performed in simulating the charged
impurity state. But in fact, we expect the procedure using the
72A-SC[Cd$^-$(2e)] to give better results that the one using the
72A-SC[Cd$^-$(fluorine)] since in the former the global properties
of the system are only smoothly affected (background density is
0.003 e/\AA$^3$ for this SC). Moreover, when considering
relaxations of atoms more distant from Cd than its NN, the
presence of fluorine atoms will spuriously influence the results,
so, in what follows we refer to the \carg.

\paragraph{Size effects}

In order to check how appropriate are the dimensions of the 72A-SC
($2a \times 2b \times 3c$) used in the present work we have
performed self-consistent electronic structure calculations for a
48A-SC ($2a \times 2b \times 2c$) and a 96A-SC ($2a \times 2b
\times 4c$). In these calculations we put the nearest oxygen
neighbors of Cd  at relative positions from Cd that correspond to
equilibrium in the 72A-SC. The size of the forces that oxygen
atoms O1 and O2 experiment in the 48A-SC and 96A-SC is a measure
of the convergence of the present calculation. We obtained that in
the 96A-SC the forces on O1 and O2 atoms point outwards (respect
to Cd) and are around \mbox{0.15 eV/\AA}. Forces of this size
produce, during the relaxation process in the 72A-SC, changes in
distances smaller than \mbox{0.01~\AA} and changes in EFG of about
\mbox{0.2 $\times$ 10$^{21}$V/m$^2$}, so, similar changes would be
expected if relaxation in the 96A-SC  was performed. In the case
of the 48A-SC forces point inwards and are of the same magnitude
 for O2 atoms but they are
about \mbox{0.65 eV/\AA} in O1 atoms. We see that there is a size
effect that makes relaxations to be larger for larger SCs, but for
the 72A-SC the effect is quite small and
 no significant variations should be expected if
relaxations in bigger SCs were considered.

\paragraph{Basis size} We compute the self-consistent electronic structure
of the 72A-SC[Cd$^-$(2e)] for the relaxed positions of
Table~\ref{tab:distances} increasing the size of the basis to 7100
LAPW functions (RK$_{MAX}$=7). Forces on O1 and O2 atoms are below
the tolerance value of \mbox{0.025 eV/\AA} indicating that the
result of the relaxation performed is unaltered for a substantial
increase of the basis size. In fact, the forces on all the other
atoms in the cell have the same values than using RK$_{MAX}$=6
(within the tolerance) except for the Ti neighbors of Cd in the
$z$ direction where the forces differ in \mbox{0.12 eV/\AA}.
Changes obtained in $V_{ii}$ components of the EFG tensor are
smaller than 0.05 $\times$ 10$^{21}$V/m$^2$.
 We have also considered the
 inclusion of a
 Cd-4$d$ LO to improve linearization. When a Cd-4$d$ LO is introduced
 with energy at the Fermi level in order to improve the description
 of the impurity states no influence in the forces is detected.
 The change in $V_{ii}$ components of the EFG is  smaller than 0.1 $\times$ \unitefg.

\paragraph{Exchange-correlation potential} We performed electronic
self-consistent calculations and relaxation of Cd nearest oxygen
neighbors for the system 72A-SC[Cd$^-$(2e)] using the generalized
gradient approximation (GGA) \cite{gga} instead of LDA. With the
use of this parametrization for the exchange-correlation potential
we obtained  2.18~{\AA} and 2.10~{\AA} for Cd-O1 and Cd-O2
distances, respectively  (i.e. only a small change of 0.01~{\AA}
in Cd-O2 distance). For the EFG we obtained
$V_{33}$=$V_{YY}$=+4.94 $\times$ \unitefg\ and $\eta$=0.37.

\paragraph{Further relaxations} In order to study the effect of relaxing the
coordinates of atoms beyond the nearest neighbors of the Cd atom
we have performed the following two additional relaxations: (a) we
allow to relax the coordinates of all atoms within a cutoff radius
 $R_C$=4~{\AA} centered at Cd (this involves 24 atoms) and
(b) idem for $R_C$=4.6~{\AA} (this involves 42 atoms). For radius
larger than 4.6~{\AA}, atoms to be relaxed would be nearer to the
images of Cd atom from neighboring cells than to the Cd itself, so
we consider this radius as a limit for the present SC. In
Table~\ref{tab:further relax} we compare the results obtained for
both relaxations with the ones corresponding to the NN relaxation.
We observe that there is not any qualitative change in the results
already discuss for the NN relaxation. There are, however, some
small variations in the values predicted for the EFG. The
differences between results from relaxations (a) and (b) (that are
as larger as the differences between relaxations (a) and NN) are
in part caused because in (b) the relaxation of O3 atom is
allowed. Atom O3 is the NN of O2 atom
 in the $X$ direction and its relaxation of about 0.04~{\AA}
 allows a further relaxation of about 0.012~{\AA}  of the O2 atom.
The fact that atom O3 is the one that experiments the
largest relaxation (not considering O1 and O2) shows
 that directional bonding plays an important role in this structure.

\begin{table}
\caption{\label{tab:further relax}Results of the different
relaxations performed for the 72A-SC[Cd$^-$(2e)]. In each case all
the coordinates of the N$_A$ atoms within a radius $R_C$ (in \AA)
are relaxed until forces on them are below 0.025 eV/\AA. All units
as in Table I and II.}
\begin{ruledtabular}
\begin{tabular}{lccccrrrr}
           & R$_C$ & N$_A$ & d(Cd-O1)  & d(Cd-O2) &
                  $V_{XX}$ & $V_{YY}$  & $V_{ZZ}$ & $\eta$  \\ \hline
NN         & 2.5   &  6    &  2.185   &  2.111    &
                  -2.87    & +4.55     &  -1.68   &  0.26    \\
(a)        & 4.0   &  24   &  2.176    &   2.104  &
                   -3.25  & +4.99     & -1.74  & 0.30         \\
(b)        & 4.6   &  42  &  2.187   &   2.116    &
                   -3.17   & +4.86   &   -1.69    &  0.30     \\
\end{tabular}
\end{ruledtabular}
\end{table}

\vskip 2mm

In summary, we observed that none of the factors considered
influence qualitatively the results, but their effect is neither
negligible. We therefore confirm our prediction about sign
(positive) and direction ($Y$)
 of $V_{33}$, but it is not
 possible to perform an exact prediction about its
 magnitude. Ours checks shows that $d$(Cd-O1) and $d$(Cd-O2)
 are converged within 0.01~{\AA} and \cgav\ and $\eta$ within 0.5 $\times$ \unitefg\
 and 0.1, respectively. Taking the values from the last row of
 Table~\ref{tab:further relax} as our predicted values for $V_{ii}$
 we obtain discrepancies with experiment:~\cite{exp2,prl}
 of 0.37 $\times$ \unitefg\ for $|V_{33}|$ and 0.12 for $\eta$ that
 could be attributed to precision errors of the present calculations.

\subsubsection{Neutral impurity: Cd$^0$}

Let us briefly discuss
 the case of the 72A-SC(Cd$^0$). Due   to   the factor four that exists because of $k$-sampling,
calculations are much more time consuming in this case. Hence, we
have checked the size of the SC performing calculations only on a
48A-SC, and also checked the size of the basis repeating the
calculation for the 48A-SC with RK$_{MAX}$=7. Comparison of the
values obtained for EFG and forces let us conclude
 that errors in the 72A-SC(Cd$^0$)  are expected to be of the
same magnitude than the ones obtained for the 72A-SC[Cd$^-$(2e)].
Only the effect of adding a Cd-4$d$ LO and the use of GGA instead
LDA produce a larger variation of the EFG in this case than in the
charged cell, and is understandable because of the larger $d$
contribution to $V_{ii}$ in the case of the neutral state of the
impurity. Calculations in the 72A-SC(Cd$^0$) with this
parametrization for the exchange-correlation potential and
introducing Cd-4$d$ LO predict values of
$V_{33}$=$V_{XX}$=-7.90$\times$ \unitefg\ and $\eta$=0.80. In
addition, we found a force of \mbox{0.04 eV/\AA} inwards at O1
atoms, so a small refinement in O1 position would be expected.
About further relaxations, we have only performed relaxation (a)
($R_C$=4~\AA) for the 72A-SC(Cd$^0$)
 and also obtain variations of the same order than in 72A-SC[Cd$^-$(2e)], in
particular we obtained $V_{33}$=$V_{XX}$=-6.8$\times$ \unitefg\
and $\eta$=0.97. Then, the same conclusions about accuracy than in
the 72A-SC[Cd$^-$(2e)] hold for the 72A-SC(Cd$^0$) but note that,
due to the high value of $\eta$ in this case, the sign and
direction of $eq$ could change because of precision.
\begin{figure*}[!]
\includegraphics*[bb= 3.1cm 7.3cm 19.5cm 22.4cm, viewport=-2cm 0cm 20cm 15cm, scale=0.85]{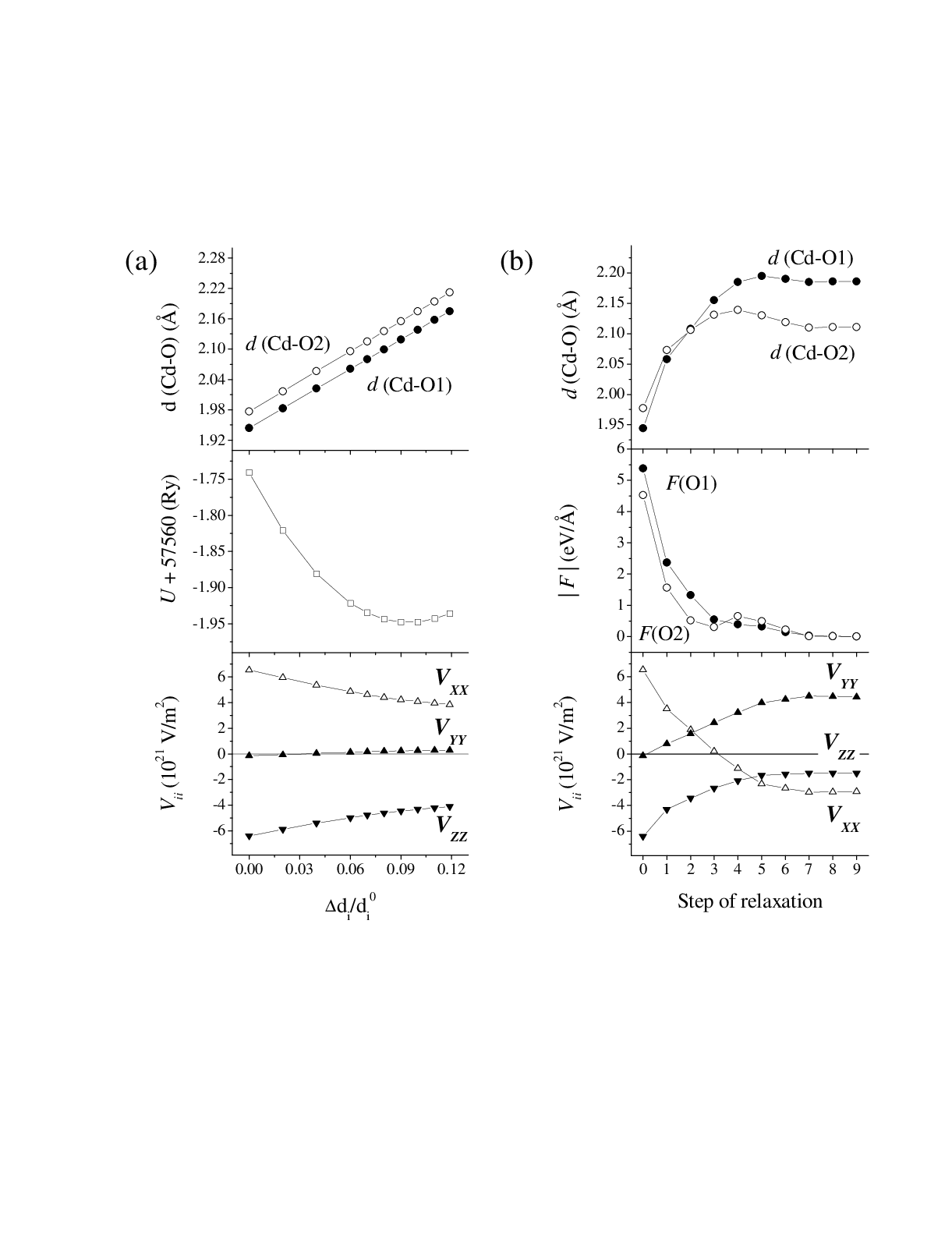}
\caption{\label{relaj}(a) Isotropic relaxation process. Distances
$d_1$=d(Cd-O$_1$) and $d_2$=d(Cd-O$_2$), total energy, and
$V_{ii}$ as a function of the relative displacement, $\Delta d_i
/d_i^0$, of the nearest O atoms for the 72A-SC[Cd$^-$(2e)] system
($\Delta d_i= d_i-d_i^0, d_i^0$ is $d_i$ for the pure system); (b)
Free relaxation process. Distances $d_1$ and $d_2$, forces at O1
and O2, and $V_{ii}$ as function of the step of the relaxation in
the 72A-SC[Cd$^-$(2e)] system.}
\end{figure*}

\subsection{\label{sec:3.E} Comparison with other calculations}

The  simplest and most widely used approximation for the
calculation of the EFG at a  probe-atom  is the point-charge model
(PCM).\cite{techniques2} In this approximation the EFG tensor at
the
 probe site is $(1-\gamma _{\infty}) V_{ij}^{latt.}$, where
$V_{ij}^{latt.}$ is the EFG tensor produced by valence nominal
charges located at the ion positions in the lattice, and $\gamma
_{\infty}$ is the Sternheimer antishielding factor\cite{stern}
that depends only on the probe-atom. In this way, this model
assumes that the symmetry and orientation of the EFG tensor at
impurity sites are unaltered by its presence.
 PCM gives for TiO$_2$(Cd): $V_{33}$=$V_{XX}$=-2.27 $\times$ \unitefg\ and
$\eta$=0.40 when a value of -29.27 is used for $\gamma
_{\infty}$.\cite{feiock}
 In Ref.~\onlinecite{lieb} the authors assume
that relaxation is responsible for the disagreement of PCM
predictions with experiment and speculate that an isotropic
relaxation of 0.04~\AA\ outwards of all the Cd nearest oxygen
neighbors would produce the desired result for $|V_{33}|$ and
$\eta$. Our results indicate that relaxations are not isotropic
and they are so much larger than this, but even if the relaxed
coordinates from our calculation (see Table I) were used, the PCM
would fail in the description of the EFG giving $V_{33}$=$V_{XX}$=
-8.20 $\times$ \unitefg\ in clear contradiction with
$V_{33}$=$V_{YY}$= +4.55 $\times$ \unitefg\ that we obtain from
the self-consistent FLAPW calculation. The disagreement of PCM
with the FLAPW prediction could not either be attributed in this
case to a change in the value of $\gamma _{\infty}$ since the sign
and directions of both predictions are different. Thus, it is
clear that the problem of EFG at cationic sites in TiO$_2$ is too
complicated to be described even approximately by simple PCM
calculations.

 In our previous work \cite{our_nqi} we performed self-consistent
electronic structure calculations of this system with a
twelve-atom super-cell (12A-SC:  $a \times b \times 2c$).
Relaxations were performed only for the neutral charge state of
the impurity and at the end of the relaxation process two
electrons were added to compute EFG without computing the
self-consistent potential of the charged cell. The
  structural relaxation obtained by this procedure were
  smaller than in the present work but account for the inversion
 of Cd-O1 and Cd-O2 distances [$d$(Cd-O1)=2.12~{\AA},
 $d$(Cd-O2)=2.07~{\AA}] with respect to the unrelaxed structure
(see Table~\ref{tab:distances}).
 The description obtained for the EFG was very
 similar to the one of the 72A-SC[Cd$^-$(2e)], but we understand that this agreement
 is somehow fortuitous because usage of 12A-SC relaxed coordinates to
  compute the EFG in the 72A-SC[Cd$^-$(2e)] gives very different results.

In Ref.~\onlinecite{akai}, the authors have used a very similar
approach as that we used  in Ref.~\onlinecite{our_nqi} using a
12A-SC, but they assumed that the relaxations of the nearest
oxygen neighbors of the Cd atom  were isotropic. As a consequence,
they obtained $d$(Cd-O1)=2.04~{\AA}, $d$(Cd-O2)=2.08~{\AA}, and
also a very different result for EFG (see Table I),
$V_{33}$=$V_{ZZ}$=-5.09 $\times$ \unitefg and  $\eta$=0.39 (for a
carefully comparison of these two calculations see
Ref.\onlinecite{our_nqi}).

A question that arises at this point is if self-consistent
electronic FLAPW calculations performed with a {\it converged} SC
give EFG compatible with experience or not in the case that
relaxations of the oxygen NN of the Cd atom are constrained to be
isotropic. This is an interesting point in order to know if
available experimental data is enough to refute the assumption
that the oxygen NN relax isotropically.
We have performed self-consistent calculations for the system
72A-SC[Cd$^-$(2e)] for different positions of O1 and O2 atoms but
moving them outwards, keeping the relation $d$(Cd-O1)/$d$(Cd-O2)
constant. We determined the equilibrium position of oxygen atoms
as the one that produce a minimum in the energy [see
Fig.~\ref{relaj}(a)]. We obtain that Cd-O distances are 2.12~{\AA}
and 2.16~{\AA} for O1 and O2 atoms, respectively [a relaxation of
9\% of the unrelaxed distances, see Fig.~\ref{relaj}(a)]. If we
compare Figs.~\ref{relaj}(a) and (b), we can note that due to the
assumption of isotropic relaxations, there is no inversion in the
Cd-O distances and, in consequence, the strong change in the EFG
components does not take place. In particular, there is not change
in sign and orientation of $V_{33}$ as in the case of our free
relaxation. At the equilibrium position we obtained for $V_{33}$ a
value of -4.46 $\times$ \unitefg\ (pointing in [001]-direction)
and a high $\eta$ value of 0.91, confirming that an isotropic
relaxation is not consistent with the experimental data.

\section{CONCLUSIONS}

In this work we have studied through a series of first-principles
calculations the problem of a Cd impurity substitutionally-located
at the cationic site in rutile TiO$_2$.
   The main result of our work, i.e. that Cd introduces in the host
   fairly anisotropic relaxations of its nearest oxygen neighbors
   and that this produces a change of orientation of V$_{33}$ from
   the [001] to the [110] direction when a Ti atom is replaced by a Cd atom in
   pure TiO$_2$, was briefly presented in a recent work with the experimental
   confirmation of the last prediction \cite{prl}.
In this work we have presented details about the electronic structure
of the different impurity systems considered.
   We have considered atomic relaxations and electronic structure
   self-consistently and have obtained that both aspects of the
   problem interact with each other.
   We obtained that atomic relaxations are different for the
   charged and neutral state of the impurity and that, on the other side,
   the relaxation process produces a drastic variation in the asymmetry
   of the charge distribution near the probe-atom for a given charge state of the
   impurity, which is detected in the strong variation
   of the EFG tensor.
We have shown that the huge difference in the values of the
asymmetry parameter $\eta$ between the charged and
 neutral state of the impurity arises because of the filling of the
 impurity level at the Fermi energy. This difference in the $\eta$ value
 determines, through comparison with experiment, that Cd is in a charged
 state when it is introduced as impurity in TiO$_2$ at room
 temperature. From these results we have confirmed that the EFG
 tensor is a very useful magnitude because it is sensible to
 subtle details of the electronic structure and it can be
 determined experimentally with high resolution.
 We have performed a series of checks of the accuracy of the present
 calculations in order to show that all the predictions of this
 work are the same if an increment in the basis size, k-mesh or
 size of the SC are considered or if a different exchange-correlation
  potential is used. We have also shown that
 considering relaxations beyond nearest neighbors
 does not produce any qualitative change in our results.
   Finally, we have checked that the hypothesis of isotropic relaxations
   and the use of PCM approximations give results incompatible with
   experiments and with our calculations. From our results it is clear that
    the problem of the EFG at Cd impurities
in TiO$_2$ is too complicated to be described (even approximately)
by simple models like PCM, antishielding factors and isotropic
relaxations.
  We can conclude that a proper theoretical
description of electronic properties of metal impurities in oxide
semiconductors should consider self-consistently the charge-state
of the impurity and the impurity-induced distortions in the host,
specially in the first shell of neighbors of the impurity.

\

\section*{ACKNOWLEDGMENTS}
We are indebted to Prof. Dr. Mariana Weissmann for fruitful
discussions and critical reading of the manuscript. We gratefully
acknowledge the support of Prof. Dr. A.G. Bibiloni on this
project. We are grateful to Dr. M. Cervera for helpful
suggestions. This work was partially supported by CONICET, Agencia
Nacional de Promoci\'on Cient\'{\i}fica y Tecnol\'ogica (ANPCyT)
under PICT98 03-03727, and Fundaci\'on Antorchas, Argentina, and
The Third World Academy of Sciences (TWAS), Italy (RGA97-057).


\begin{thebibliography}{}
\bibitem{dnd} When axial symmetry is present (this is the case of the
variation in the $d$ contribution that we try to explain) it is
useful to define the asymmetry count, which for $d$ orbitals is:
$\Delta n_d = n_{d_{yz}} - 1/2 (n_{d_{xy}}+n_{d_{xz}})
            + 1/2 (n_{d_{z^2}} - n_{d_{x^2-y^2}})$,  where $n_i$ is the
charge of $i$-character (in electrons) inside the MT sphere and $x$ is the axial
axis. With this definition $V_{XX}(d)$=$I_d \Delta n_d$ and
$V_{YY}$=$V_{ZZ}$=-$V_{XX}$/2. Note that this way of defining
 $\Delta n_d$ differs from the usual one in FLAPW \cite{efg}
 only because of the choice of the axial axis.
\end{thebibliography}
\end{document}